\documentclass[prl,twocolumn,showpacs,preprintnumbers,amsmath,amssymb]{revtex4}

\usepackage{graphicx}
\usepackage{bbm}
\usepackage{float}

\begin{document}

\preprint{}
\title{Magnetic-induced phonon anisotropy in ZnCr$_2$O$_4$ from first principles}
\author{Craig J. Fennie and Karin M. Rabe}
\affiliation{Department of Physics and Astronomy, Rutgers University,
        Piscataway, NJ 08854-8019}
\date{\today}

\begin{abstract}
We have studied the influence of magnetic order on the optical phonons
of the geometrically frustrated spinel ZnCr$_2$O$_4$ from first-principles.
By mapping the first-principles phonon calculations onto a Heisenberg-like 
model, we developed a method to calculate exchange derivatives and 
subsequently the spin-phonon couping parameter from first-principles.
All calculations were performed within LSDA+U.

\end{abstract}

\pacs{63.20.-e, 75.10.Hk, 75.50.Ee, 75.30.Et}

\maketitle

%%%%%%%%%%%%%%%%%%%%%%%%%%%%%%%%%%%%%%%%%%%%%%%%%%%%%%%%%%%%%%%%%%%%%%%%%%%%%%%%%%
%INTRODUCTION
%%%%%%%%%%%%%%%%%%%%%%%%%%%%%%%%%%%%%%%%%%%%%%%%%%%%%%%%%%%%%%%%%%%%%%%%%%%%%%%%%%
%\section{Introduction}
%\label{sec:intro}

The interplay between spin and lattice degrees of freedom give rise to
an amazing variety of different phenomena, including the 
spin-Teller~\cite{oleg.prb.02} and the magnetoelectric~\cite{fiebig.review} 
effects. Perhaps the simplest example of a spin-lattice coupling is the
influence of magnetic order on phonons. Though first-principles density 
functional methods have been highly successful in describing the structural 
and magnetic properties of dielectrics~\cite{waghmare.book.05}, little 
has been done to address the coupling of spins and phonons. In this letter we 
present for the first time an approach to predict the influence of magnetic
order on optical phonons from first-principles and apply this approach to the strongly 
geometrically frustrated spinel zinc chromite. Our method is general and
has also been recently applied to provide a new understanding of the phonon
anomalies observed at the ferromagnetic (FM) transition in the chalcogenide 
spinels~\cite{fennie.spinels.06}.

Zinc chromite, ZnCr$_2$O$_4$, crystallizes in a spinel structure. 
The Cr$^{3+}$ ions (S= 3/2) form a network of vertex-sharing tetrahedra 
(a pyrochlore lattice) with strong antiferromagnetic (AFM)
interactions between nearest neighbor spins, $\Theta_{CW}$$\approx$ 
400K. As a result, zinc chromite is strongly geometrically frustrated
as evidenced by the rather low T$_N$ = 12.5K (compared to $\Theta_{CW}$).
The physics of ZnCr$_2$O$_4$ at T$_N$ involves a first-order cubic-to-tetragonal
(c-t) structural transition (to relieve geometric frustration)  as 
it enters the Neel state, bypassing a bond-ordered state~\cite{oleg.prb.02}.
Recently, Sushkov et al$.$~\cite{sushkov.prl.05} measured the 
reflectivity spectrum of ZnCr$_2$O$_4$ and found a large infrared-active
(i.r$.$) phonon splitting (11 cm$^{-1}$) at T$_N$. They invoke 
a couling between phonons and spins to argue that i.r$.$ phonons
provide a quantitative measure of the spin-Peierls order parameter~\cite{oleg.prb.02}.
Yet, a priori, it is unclear whether this spin-phonon mechanism
can account for the size of the effect.

We study the influence of magnetic order on the phonon frequencies of 
ZnCr$_2$O$_4$ from first-principles. We give evidence that anisotropy
induced by AFM ordering of spins can account for the large anisotropy 
measured in the i.r$.$ phonons~\cite{sushkov.prl.05}. We decouple the 
macroscopic elastic degrees of freedom from those of the spins by
calculating the phonons of crystallographically cubic ZnCr$_2$O$_4$.
This approach is similar to that taken by Massidda et al$.$~\cite{massidda.prl.99} 
for MnO and naturally decouples the tetragonality induced by the 
lattice strain at T$_N$ from that generated by the spin pattern
alone. 
Next, in order to understand the first-principles results, we revisit the 
long-standing problem of how spins
couple to optical phonons~\cite{baltensperger}:
$$\omega\approx\omega_0+\lambda \langle{\bf S}_i\cdot{\bf S}_j \rangle $$
We develop an approach where for the first time experimentally accessible
quantities of this theory can be calculated from first-principles. 
Finally, we apply this method to calculate the
coupling parameter, $\lambda$, for ZnCr$_2$O$_4$.  We find it
to be in excellent agreement with that measured, providing
confirmation of a strong spin-phonon mechanism responsible for the
large phonon splitting.

%%%%%%%%%%%%%%%%%%%%%%%%%%%%%%%%%%%%%%%%%%%%%%%%%%%%%%%%%%%%%%%%%%%%%%%%%%%%%%%%%%
% METHOD
%%%%%%%%%%%%%%%%%%%%%%%%%%%%%%%%%%%%%%%%%%%%%%%%%%%%%%%%%%%%%%%%%%%%%%%%%%%%%%%%%%

First-principles density-functional calculations using projector 
augmented-wave potentials were performed within LSDA+U (local 
spin-density approximation plus Hubbard U)~\cite{anisimov.jpcm.97} 
as implemented in the {\it Vienna ab initio Simulation Package}~\cite{VASP,PAW} 
with a plane wave cutoff of 500 eV and a 6$\times$6$\times$6 $\Gamma$-centered 
$k$-point mesh. All calculations were performed with collinear spins and without
LS-coupling (inclusion
of LS-coupling does not change the results). We performed full optimization
of the lattice parameter, $a$= 8.26 \AA\ (exp: $a$=8.31 \AA), and 
anion parameter, $u$ = 0.386 (exp: $u$=0.387), in space group Fd$\bar{3}$m, 
where we find excellent agreement with experiment. The Cr on-site Coulomb, 
U = 3 eV, and exchange, j=0.9 eV, parameters are justified {\it a posteriori} 
from the calculation of the magnetic exchange constant as we now 
discuss~\cite{fennie.prb.05c}.

For zinc chromite, the nearest neighbor (n.n$.$) exchange constant, $J$, 
is an order of magnitude greater than all next nearest neighbor 
interactions. $J$ is determined by a balance between AFM direct Cr-Cr 
exchange and FM 90$^{\circ}$ Cr-O-Cr superexchange and to a large extent, 
which one wins can be attributed to volume and electronegativity of 
the anion.~\cite{goodenough} Exchange constants can be extracted by
mapping first-principles calculations of the total energy for different
spin configurations at T=0 onto a classical Heisenberg model; 
$E_{spin} = $$- 2J\sum_{\langle nn \rangle} {\bf S}_i\cdot{\bf S}_j $
(sum over $z$=6 nn's).  This gives for the energy of the primitive
unit cell (4 Cr-ions): $E_{FM}$=$E_0 - 24JS^2$ and $E_{AFM}$=$E_0 + 8JS^2$. 
From this we calculate J = -2.1 meV which compares well with J$_{exp}$ = -2.25 meV 
estimated from the measured $\Theta_{CW}$. %Within LSDA, J is twice as
%large, highlighting the need to include correlations such as within LSDA+U. 

{\it First-Principles Phonons.}$-$For cubic ZnCr$_2$O$_4$, group-theoretical 
analysis predicts that 
the 39 zone-center optic modes transform according to the following
 irreducible representations (irreps) of $O_h$: $\Gamma_{i.r.}= 4 T_{1u}$,
$\Gamma_{Raman}= 3 T_{2g} \oplus 1 E_g \oplus 1 A_{1g}$, and
$\Gamma_{silent}= 2 A_{2u}\oplus  2 E_u \oplus 2 T_{2u} \oplus 1 T_{1g}$,
and the three acoustic modes according to $\Gamma_{acoustic}= 1 T_{1u}$. 
With AFM spins, the ordering pattern (the ``collinear model'' of Ref.~\onlinecite{oleg.prb.02})
induces a tetragonal axis chosen here to be $\hat{z}$.
Under this tetragonal
"distortion", the point group is lowered from $O_h\rightarrow D_{4h}$, 
and the irreps previously discussed become reducible, although the Raman
and i.r$.$-active irreps still do not mix.  Therefore we can continue to
talk about these modes independently  and although the Raman modes
do split, this splitting is small and not our main concern. Still, to 
demonstrate the quality of the phonon calculations for zinc chromite, 
we present our calculated Raman-active frequencies (cm$^{-1}$): $T_{2g}$:
185 (186), 521 (515), 608 (610); $E_{g}$: 466 (457); $A_{1g}$ 687 (692), 
which are in remarkable agreement ($\sim$1$\%$) with experiment (shown in 
parenthesis)~\cite{himmrich.ssc.79}.

\begin{figure}[t]
\includegraphics[width=7.5cm,height=4.5cm]{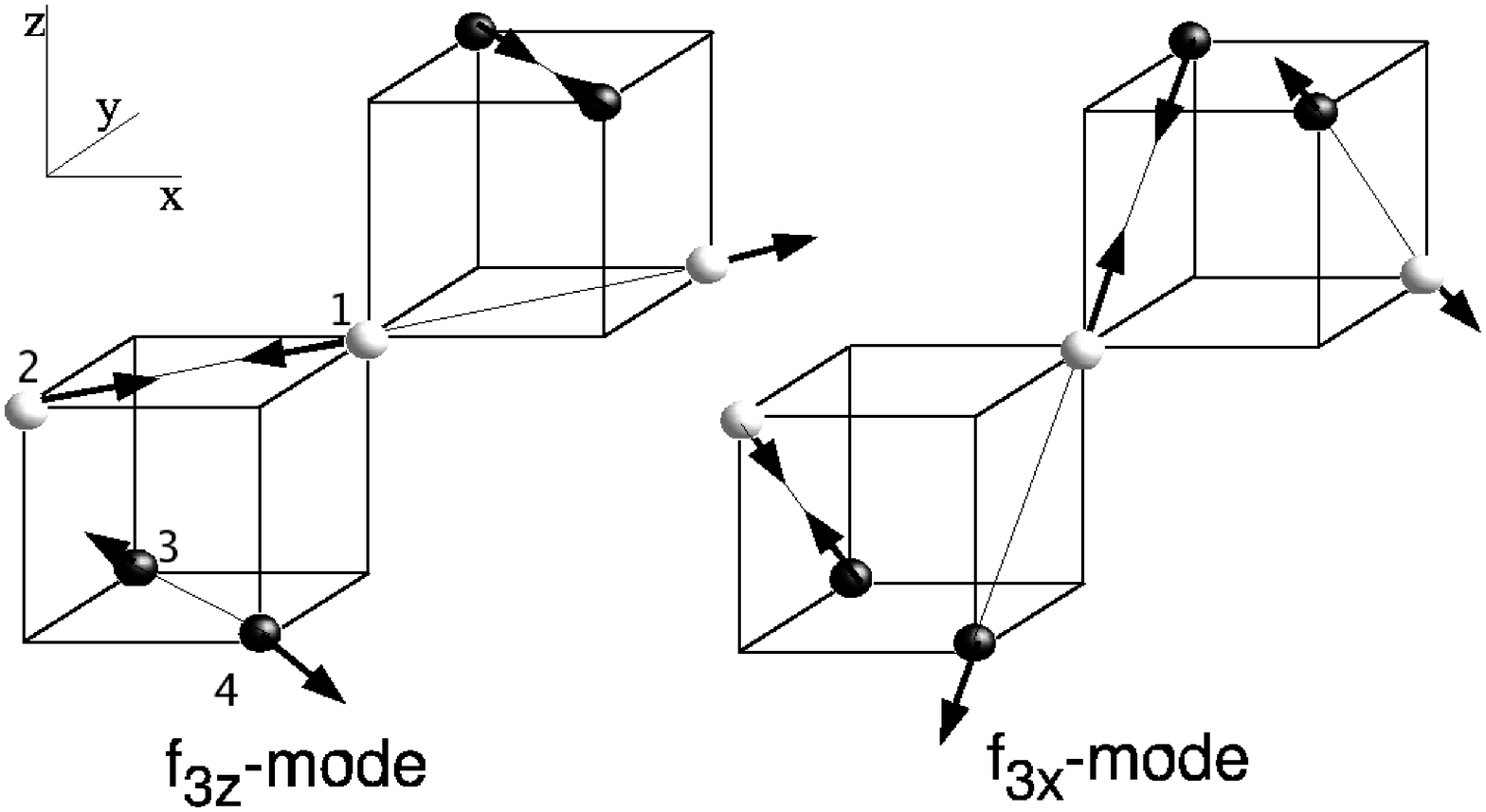}\\
\caption{\label{fig:FIG1} 
Partner functions $f_{3z}$ (left) and $f_{3x}$ (right) that 
transform like different rows of irrep T$_{1u}$. Cr-up 
spins (white spheres) and Cr-down spins (black spheres) are 
shown while oxygen atoms sit at the unoccupied corners are not 
shown for clarity. Arrows indicate direction 
of atomic displacements consistent with $f_{3\alpha}$.}
\end{figure}

We now proceed 
to the main part of this letter, the influence of magnetic order on the 
i.r$.$ phonons. Under $D_{4h}$ the i.r$.$ modes split according to 
$T_{1u}$$\rightarrow$$A_{2u}$$\oplus$$E_u$, where the 1-d $A_{2u}$ and 
2-d $E_u$ irreps transform like vectors along and perpendicular to the
tetragonal axis respectively. Additionally, certain rows of the silent 
$T_{2u}$ irrep become i.r$.$-active and can in principle mix with those
originating from $T_{1u}$. This mixing is very small and for our purposes
acceptable to ignore. For the remainder of this letter we only discuss the 
4 modes originating from $T_{1u}$. The rows of $T_{1u}$ do not mix under
$D_{4h}$ so that we can still talk about the $x$, $y$, ($E_u$) and $z$ 
($A_{2u}$) blocks of the dynamical matrix. We compute the i.r$.$ phonons by 
constructing the three 5x5 blocks (4 i.r$.$-active, 1 acoustic) of the 
dynamical matrix from Hellmann-Feynman forces on the symmetry-adapted 
partner functions, $f_{n,\alpha}$, where $n$=1,5 and $\alpha$= x,y,z label
the partner function and the row of T$_{1u}$, respectively. These partner 
functions transform like vectors along $\hat{\alpha}$ under O$_h$. The partner
functions, $f_{1,\alpha}$, $f_{2,\alpha}$, and $f_{4,\alpha}$, involve atomic 
displacements along $\hat{\alpha}$ of the entire zinc, chromium, and oxygen 
sublattices, respectively. Nontrivial displacements within the chromium
and oxygen sublattices, whose partner functions we label $f_3$ and $f_5$ 
respectively, are also possible. Although $f_{3,\alpha}$ and $f_{5,\alpha}$ 
both transform like vectors along $\hat{\alpha}$, atomic displacements 
associated with these partner functions take place in the plane perpendicular
to the $\alpha$-axis. In Fig.~\ref{fig:FIG1} we show $f_{3,z}$ and
$f_{3,x}$ which are seen to strongly modulate the Cr-Cr bond length.
Their importance to the present problem will soon become apparent.

%%%%%%%%%%%%%%%%%%%%%%%%%
%%%%%%%%%%%%%%%%%%%%%%%%%%

Our calculated i.r$.$ phonons for crystallographically cubic ZnCr$_2$O$_4$ 
with AFM are shown in Table~\ref{table:ir} labeled {\it ab initio}. We see
that AFM order induces a large tetragonal anisotropy in the phonon sector
($\omega_x$=$\omega_y$$\neq$$\omega_z$) with large splitting
($\Delta\omega$$\equiv$$\omega_z$-$\omega_x$) occurring for modes (1) and
(2), $\Delta\omega$ $\approx$ 0.12$\omega_z$, and smaller spitting for (3)
and (4), $\Delta\omega$ = 0.02$\omega_z$. (we hold off for now on comparing
these results to experiment). 
This mode dependent behavior is reminiscent of the phonon anomalies displayed 
at T$_c$ in the FM spinel CdCr$_2$S$_4$~\cite{wakamura.jap.88} where 
insight was gained by looking at the eigendisplacements. We don't repeat
the analysis hear except to make two points. First, all four modes 
contain Cr-$f_3$ partner function (more on this below) so
that the Cr-Cr bond is highly distorted in the plane perpendicular to 
$\alpha$ but in  (1) and (2) the total distortion is such that all Cr-O 
bond lengths remain fixed while in  (3) and (4) the Cr-O bond lengths, 
particularly along $\hat{\alpha}$, are strongly modulated (we will see 
what this implies). Second, the presence or absence of Zn-$f_1$ has no 
effect on the spin-phonon coupling in contrast to what is commonly 
proposed. We understand this from our calculations of the $q$=0 component
of the force constant matrix, $C_{n,\alpha;n',\alpha'}$, where we find
$C_{1,x;1,x}$ = $C_{1,z;1,z}$, as you would expect for cubic symmetry. 
We found similar results in CdCr$_2$S$_4$, 
CdCr$_2$Se$_4$, and HgCr$_2$Se$_4$ which suggests that the previous
interpretation of phonon anomalies in these FM spinels should be 
revisted~\cite{fennie.spinels.06}.

Our first-principles calculations clearly show that AFM spin-ordering induces 
a large i.r$.$ phonon 
splitting as shown in Table~\ref{table:ir}. Sushkov argued that the large 
phonon splitting of $T_{1u}$(2) was caused through modulation of direct Cr-Cr 
exchange by $f_{3}$, and since $T_{1u}$(2) contained the largest amount of 
this partner function, it was the only splitting measurable. In Fig.~\ref{fig:FIG1} 
the anisotropy imposed by the magnetic order on this mode is clear. Supporting
their argument is our calculations of the $q$=0 component of the force constant
matrix. Looking at the diagonal components we find that the anisotropy, 
$C_{n,z;n,z}$-$C_{n,x;n,x} $ (which equals zero for a cubic system), is 
$\approx$20$\%$ for $f_3$, being 3 times larger than that due to any other 
partner function. Second, T$_{1u}$(2) contains twice as much $f_3$ compared 
with the other i.r$.$ modes. But in contrast to what is seen experimentally, 
the phonon frequencies of all four modes are split. Additionally, the calculated
splitting of $T_{1u}$(2), $\Delta\omega$=50 cm$^{-1}$, is about 4.5 times larger
than the measured splitting, $\Delta\omega$=11 cm$^{-1}$. One question then 
becomes how do we interpret our phonon calculations to be consistent with experiment? 
Also what do they teach us about the spin-phonon mechanism and the role of 
direct exchange? Clearly a more systematic approach to understand the first-principles
calculations is desirable.

%%%%%%%%%%%%%%%%%%%%%%%%%%%%%%%%%%%%%%%%%
%SECTION TABLES: PHONONS
\begin{table}[t]
\caption{ Infrared-active, T$_{1u}$, phonons frequencies, cm$^{-1}$, of ZnCr$_2$O$_4$.}
\begin{ruledtabular}
\begin{tabular}{lccccccccccc}
& \multicolumn{2}{c}{Experiment } 
& &\multicolumn{8}{c}{Present Theory} \\ % \cline{4-12}
& \multicolumn{2}{c}{Ref.~\onlinecite{sushkov.prl.05}}  
& \multicolumn{4}{c}{Model}&&&&\multicolumn{2}{l}{ {\it ab intio}} \\ \cline{4-7}  \cline{11-12}

\begin{tabular}{l} \\  \\(1)\\(2)\\(3)\\(4) \\  \end{tabular}

&\begin{tabular}{c} 13K  \\ \\ 186\\371\\501\\619\\  \end{tabular}
&\begin{tabular}{cc} \multicolumn{2}{c}{9K } 
\\$\hat{x}$&$\hat{z}$ \\\multicolumn{2}{c}{186+$\delta$} \\368&379\\
\multicolumn{2}{c}{501}\\\multicolumn{2}{c}{619}\\  \end{tabular}

&
\begin{tabular}{c} PM\\ \\189 \\366\\514\\621\\   \end{tabular}
&\begin{tabular}{cc} \multicolumn{2}{c}{AFM}\\ $\hat{x}$&$\hat{z}$
\\186&191\\361&372\\514&517\\621&623\\   \end{tabular}
&\begin{tabular}{r} $\mathcal{J}''_\perp$\\ \\-240 \\-999\\-254\\-66\\   \end{tabular}
&\begin{tabular}{r} $\mathcal{J}''_\|$\\ \\0 \\22\\225\\257\\   \end{tabular}
&
&&&
\begin{tabular}{cc} \multicolumn{2}{c}{AFM}\\ $\hat{x}$&$\hat{z}$
\\174&198\\342&392\\510&526\\620&630\\   \end{tabular}

%\footnotetext[1]{Ref.~\onlinecite{himmrich.ssc.79}}
%\footnotetext[2]{Ref.~\onlinecite{sushkov.prl.05}}
%\footnotetext[3]{T=0; from model, see text}

\end{tabular}
\end{ruledtabular}
\label{table:ir}
\end{table}

%%%%%%%%%%%%%%%%%%%%%%%%%%%%%%%%%%%%%%%%%

%%%%%%%%%%%%%%%%%%%%%%%%%%%%%%%%%5

{\it First-Principles Model}.$-$We begin by writing the total energy, 
$E$=$E_0$+$E_{ph}$+$E_{spin}$, of a system of i.r$.$ phonons, 
$E_{ph}$=$ \frac{1}{2} \sum_{\eta \eta'} C_{\eta,\eta'}
f_{\eta} f_{\eta'}$,  $\eta$$\equiv$$\{n,\alpha\}$ and 
$C_{\eta,\eta'}$=$\delta_{\alpha\alpha'}$$C_{n,n'}$, and spins, 
$E_{spin}$=$-\sum_{ij} J_{ij}{\bf S}_i\cdot{\bf S}_j $. Without 
the spin-orbit interaction the spin and phonon sectors are still indirectly 
coupled due to a positional dependence of the exchange interactions~\cite{baltensperger}.
This dependence is rather
complicated because in addition to the positions of the two spins, the
positions of the surrounding anions are important for the superexchange
process, and may even influence direct exchange~\cite{goodenough}. In 
principle $J_{ij}$ depends on the positions of {\it all} $N$ magnetic 
and non-magnetic ions, $J_{ij}(r_1,r_2, ...r_N)$. A solution then
becomes clearer if we write $J$ in terms of $f_{\eta}$ since these functions
provide a complete basis to represent all possible i.r$.$ atomic motion. 
For small distortions, $J$ can be expanded in $|f_{\eta}|$ leading to
%$\tilde{C}_{\eta,\eta'}$ = $ C_{\eta,\eta'} - \sum_{ij} {\partial^2 J_{ij}
%\over \partial f_{\eta} \partial f_{\eta'}} \langle{\bf S}_i\cdot{\bf S}_j \rangle$
$$\tilde{C}_{\eta,\eta'} =  C_{\eta,\eta'} - \sum_{ij} {\partial^2 J_{ij}
\over \partial f_{\eta} \partial f_{\eta'}} \langle{\bf S}_i\cdot{\bf S}_j \rangle$$
to lowest order in $S_i$$\cdot$$S_j$.
In general, symmetry of the magnetic system can reduce this expression further. 
For spinels, $D_{4h}$ does not induce couplings between different rows of 
$O_h$, i.e$.$ $\tilde{C}_{\eta,\eta'}$=$\tilde{C}_{n\alpha,n'\alpha}$$\equiv$ 
$\tilde{C}_{n,n'}(\alpha)$, so in fact we only need to calculate 
derivatives of the type, $\partial^2 J/\partial f_{n} \partial f_{n'}$ 
(magnetic order other than $D_{4h}$ could introduce derivatives connecting
different rows of $O_h$ but will not change the value of the intra-row type). 
Now since the crystallographic group is cubic there are only two distinct 
changes in $J_{ij}$ depending on whether the spins lie in a plane 
perpendicular to $\hat{\alpha}$ or not. With this let us define
${\partial^2 J_{ij}/\partial f_{\eta} \partial f_{\eta'}} \equiv J''_{\perp n,n'}
\neq 0 \,\,\,\, \forall \,\,\,\, \hat{r}_{ij} \cdot \alpha=0$
and
${\partial^2 J_{ij}/\partial f_{\eta} \partial f_{\eta'}} \equiv J_{\| n,n'}''
\neq 0 \,\,\,\, \forall \,\,\,\, \hat{r}_{ij} \cdot \alpha \neq0$
where $ \hat{r}_{ij}$ is the direction vector linking nearest neighbor spins, 
$S_i$ and $S_j$. Then we can write
%
%$\tilde{C}_{n,n'}(\alpha)$ = $C_{n,n'}(\alpha)$-$J''_{\perp n,n'} \sum_{\hat{r}_{\perp}}
%\langle{\bf S}_i\cdot{\bf S}_j \rangle$-$J''_{\| n,n'} \sum_{\hat{r}_{\|}}
%\langle{\bf S}_i\cdot{\bf S}_j \rangle$
$$\tilde{C}_{n,n'}(\alpha) = C_{n,n'}(\alpha)-J''_{\perp n,n'} \sum_{\hat{r}_{\perp}}
\langle{\bf S}_i\cdot{\bf S}_j \rangle-J''_{\| n,n'} \sum_{\hat{r}_{\|}}
\langle{\bf S}_i\cdot{\bf S}_j \rangle$$
with the first sum over the two neighbors in a plane perpendicular to $\hat{\alpha}$ 
and the second over the remaining four.

We have considerably reduced the problem by writing the exchange 
constants in terms of the partner functions and by applying symmetry, 
but we are still faced with the challenge to calculate $J''$. For this 
we propose a procedure that is similar to the approach of calculating
exchange constants - by mapping onto a Heisenberg-like model. 
As such let us write
\begin{eqnarray}
\tilde{C}_{n,n'}^{FM}(\hat{z})& =& C_{n,n'}^{PM} 
- 8 J''_{\perp n,n'} S^2 -16 J''_{\| n,n'} S^2  \nonumber \\
\tilde{C}_{n,n'}^{AFM}(\hat{z})&=& C_{n,n'}^{PM} 
- 8 J_{\perp n,n'}'' S^2 + 16 J_{\| n,n'}'' S^2   \nonumber \\
\tilde{C}_{n,n'}^{AFM}(\hat{x})&=& C_{n,n'}^{PM} + 8 J_{\perp n,n'}'' S^2  
\nonumber \\ \nonumber
\end{eqnarray}
where $\tilde{C}^{FM}$ and $\tilde{C}^{AFM}$ are the force constant matrices 
calculated from first-principles with FM and AFM order respectively. Here we
have three equations and three unknowns. In Table~\ref{table:ir} we show the
$T$=0 PM phonon frequencies extracted from the model which are found to 
be in excellent agreement (1-2$\%$) with those measured at 13K. To test the
validity of the model w.r.t$.$ first-principles we perform additional phonon 
calculations with different magnetic order, for example, ferrimagnetic (FiM) 
order with one Cr-ion spin-down, the other three spin-up. For this spin 
configuration according to our model $\tilde{C}^{FiM}(x)$=$\tilde{C}^{FiM}(y)$
=$\tilde{C}^{FiM}(z)$=$C^{PM}$. If we now perform first-principles calculations
of the phonons with FiM magnetic order, we find that the frequencies 
are in exact agreement ($<$1cm$^{-1}$) with those we extracted from the model, i.e$.$ 
$\omega_{\lambda}^{FiM}$= $\omega_{\lambda}^{PM}$~\cite{comment.model}.
As a further test we considered additional FiM spin configurations within
the conventional unit cell (eight formula units). The model 
reproduced the first-principles calculations to within 1 cm$^{-1}$ in every case.

{\it Discussion}.$-$Having established the model correctly captures the 
influence of magnetic order on the first-principles phonons, we turn our
attention to calculating experimentally accessible quantities of the problem.
First, on the role of direct exchange, it is telling that we 
find $\partial^2 J_{\perp}/\partial f_3 \partial f_3$
$\approx$-0.172 meV/\AA$^2$, more than $\sim$10 times larger 
than any other component of $J_{\perp}''$ or  $J_{\|}''$. 
This anomalously large value is consistent with the anisotropy
in the force constant matrix and, as we now show, assumed correctly 
by Sushkov to originate from the modulation of direct exchange, $J_d$. 
For small displacements $J_d$ goes like $J_d(R_c)$$\approx$$J_d e^{-\alpha\Delta R_c}$, 
where $R_c$ is the distance between neighboring Cr-atoms. 
Assuming the contribution from superexchange (SE) is small, we 
find $\alpha$=9.05\AA$^{-1}$\ which compares well 
with the measured value of $\alpha$=8.9\AA$^{-1}$. This
must be considered somewhat fortuitous since it is not clear that either
the theoretical or the experimental approach isolates direct exchange. What does
suggest the dominate role of direct exchange in $J''_{\perp 3,3}$ are our calculations
for CdCr$_2$O$_4$. This compound has a larger lattice constant and less negative
$J$ than ZnCr$_2$O$_4$ so that direct exchange is expected to be weaker.
Repeating the phonon calculations and mapping procedure for CdCr$_2$O$_4$,
we find $\partial^2 J_{\perp}/\partial f_3 \partial f_3$ $\approx$ -0.09
meV/\AA$^2$.  Considering a small contribution from SE is 
expected this compares well with $\alpha^2$J=-0.07 meV/\AA$^2$. Furthermore, 
no other components of $J_{\perp}''$ or  $J_{\|}''$ %for CdCr$_2$O$_4$
change appreciatively from those of ZnCr$_2$O$_4$, consistent with the fact that
the only partner function to change $R_c$ is $f_3$.

\begin{figure}[t]
\includegraphics[width=7.5cm,height=4.4cm]{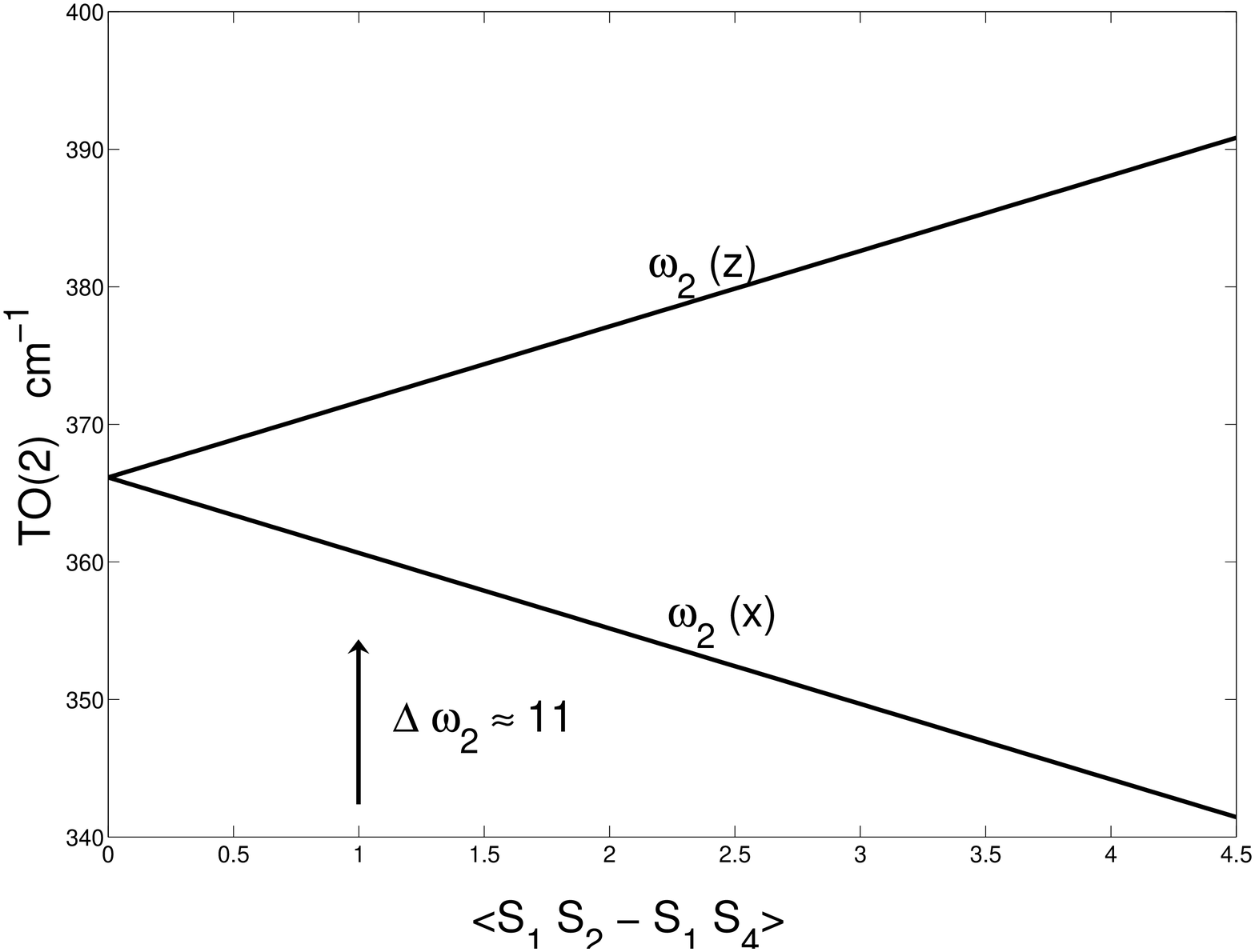}\\
\caption{\label{fig:FIG2} 
Phonon frequency of $\omega_2(\hat{z})$ and $\omega_2(\hat{x})$ as a function
of the Spin-Peierls order parameter $\langle{\bf S}_1\cdot{\bf S}_2
- {\bf S}_1\cdot{\bf S}_4\rangle $.}
\end{figure}

The relation between $J_{\perp 3,3}''$ and direct exchange being
established, we return to the discussion of the phonon splitting 
in ZnCr$_2$O$_4$. To quantify the mode dependence and compare with
experiment, let $\mathcal{J}_{\lambda}''$=$u_{\lambda}^{\dagger}$$J''$$u_{\lambda}$, 
where $u_{\lambda}$ are the PM dynamical matrix 
real-space eigenvectors. Then we can write for the phonon anisotropy 
$$\omega_{\lambda}(\hat{z})-\omega_{\lambda}(\hat{x}) \approx
4  \left( { \mathcal{J}_{\lambda \|}''- \mathcal{J}_{\lambda \perp}''  
\over \omega_{PM} } \right)\langle{\bf S}_1\cdot{\bf S}_2 - {\bf S}_1\cdot{\bf S}_4\rangle$$
where  $\langle{\bf S}_1\cdot{\bf S}_2 - {\bf S}_1\cdot{\bf S}_4\rangle$ is
defined as the spin-Peierls order parameter, $f_{sp}$,~\cite{oleg.prb.02}.
In Table~\ref{table:ir}, we see that the anisotropy experienced by $T_{1u}$ 
(1) and (2) is due to $\mathcal{J}_{\perp}''$, which is dominated by $J_{\perp 3,3}''$
and subsequently direct exchange. In contrast, $\mathcal{J}_{\|}''$ plays a
large role in the anisotropy of $T_{1u}$ (3) and (4). This is due to the 
strong modulation of the Cr-O bonds which we suspect modifies FM superexchange
and may also explain the sign. We can now compute the spin-phonon coupling 
constant, $\lambda$, for $T_{1u}$(2).  We find $\lambda_2$=${4 (\mathcal{J}_{\|}''
-\mathcal{J}_{\perp}'') /  \omega_{PM}}$$\approx$11 cm$^{-1}$ which agrees 
well with experiment, $\lambda_2$= 6-10 cm$^{-1}$. {\it So from calculations 
of just the phonons for selected spin configurations, we are able to extract
the spin-phonon coupling parameter, $\lambda$, from first-principles}. It is
also now clear that the calculated splittings are significantly larger than 
experiment because the spin-Peierls OP for collinear AFM, $f_{sp}$ = 4.5, is 
significantly larger than experiments (extracted values range from $f_{sp}$ = 1.1 to 1.8).
In Fig.~\ref{fig:FIG2} we show how the splitting for $T_{1u}$(2) evolves with
the spin-Peierls OP, where at $f_{sp}$ = 1 we recover the experimentally observed
splitting. If we substitute this value of the spin-Peierls OP into our model, we 
find that the splittings of the other modes are significantly reduced (2-5 cm$^{-1}$) 
as shown in Table~\ref{table:ir}. Although additional effects of the structural 
transition at T$_N$ still need to be worked out, perhaps such small splittings 
are not easily measurable. But, it is clear, if ZnCr$_2$O$_4$ could be prepared in a state
where $f_{sp}$ attained its maximum value the splitting would be 2.5 to 4 times greater 
than that currently measured~\cite{sushkov.prl.05}.

%%%%%%%%%%%%%%%%%%%%%%%%%
%%%%%%%%%%%%%%%%%%%%%%%%%%

In summary, we have developed a method to calculate parameters of a spin-phonon theory
from first-principles. The method
accounts for direct exchange which was shown to be responsible for the
large magnetic induced anisotropy in the phonon channel of AFM 
ZnCr$_2$O$_4$ but also equally for superexchange. As a consequence
a natural solution to the problem of phonon anomalies in FM spinels
is also revealed as a balance between these two processes.~\cite{fennie.spinels.06}
We anticipate that the approach developed will also find application 
to magnetoelectric problems where spin fluctuations coupled to optical
phonons have been proposed to explain the magnetocapacitance effect. 
Given the breath of such problems today, how the spins couple to the 
lattice and an approach to calculate the relevant parameters of such 
a theory from first-principles are important questions that needed to be addressed.

%\acknowledgments
Useful discussions with P. Chandra, S-W$.$ Cheong, M.H. Cohen, H.D. Drew, P. Mehta,
A.B. Sushkov, P. Sun and  D.H$.$ Vanderbilt 
are acknowledged. 
CF acknowledges support of Bell Labs and Rutgers University.

%%%%%%%%%%%%%%%%%%%%%%%%%%%%%%%%%%%%%%%%%%%%%%%%%%%%%%%%%%%%%%%%%%%%%%%%%%%%%%%%
%BIBLIO
%%%%%%%%%%%%%%%%%%%%%%%%%%%%%%%%%%%%%%%%%%%%%%%%%%%%%%%%%%%%%%%%%%%%%%%%%%%%%%%

\newpage


\begin{thebibliography}{10}
\bibitem{comment.model}
This isn't completely correct. The symmetry induced by FiM1 is $D_{3h}$, 
and components of the dynamical matrix connecting different rows of the irreps
of $O_h$ are not zero. The first-principles phonons were calculated by
projecting into the $T_{1u}$ subspace which has the effect of zero-ing these
"off-diagonal" terms. If one wanted to capture this physics we must include
terms like $\partial^2 J/\partial f_{n,\alpha} \partial f_{n',\alpha'}$ with
 $\alpha \neq \alpha'$ which are not important in the present problem.
\end{thebibliography}
\end{document}